\documentclass[12pt, onecolumn, amsmath, amssymb, a4paper]{aastex631}

\usepackage{amsmath, bm}
\usepackage{amssymb}
\usepackage{graphicx}
\usepackage{xcolor}
\usepackage[final]{microtype}
\def\apjl{ApJL}               % Astrophysical Journal, Letters
\def\apjs{ApJS}               % Astrophysical Journal, Supplement
\def\aap{A\&A}                % Astronomy and Astrophysics
\begin{document}
\title[]{Extreme mass ratio inspirals}
\author[0000-0001-9880-8929]{Andrea Derdzinski}
\affiliation{Department of Life and Physical Sciences, Fisk University, 1000 17th Avenue N., Nashville, TN 37208, USA}
\affiliation{Department of Physics \& Astronomy, Vanderbilt University,
2301 Vanderbilt Place, Nashville, TN 37235, USA}
\email{aderdzinski@fisk.edu}
\author{Lorenz Zwick}
\affiliation{Center for Theoretical Astrophysics and Cosmology, Institute for Computational Science, University of Zurich, Winterthurerstrasse 190, CH-8057 Zurich,
Switzerland}
\affiliation{Niels Bohr International Academy, Niels Bohr Institute, Blegdamsvej 17, 2100 Copenhagen, Denmark}
\email{lorenz.zwick@gmail.com}

\begin{abstract}
This text will appear as Section II of Chapter 5 of the book \emph{Black Holes in the Era of Gravitational-Wave Astronomy}. As a stand alone text, it serves as a brief overview of astrophysics and gravitational wave radiation of extreme mass ratio inspirals, or EMRIs. Topics covered consist of: dynamical and gas-assisted formation channels, basics of EMRI dynamics and gravitational radiation, and science potential for both astrophysics and fundamental physics.
\end{abstract}

\section{Introduction}
The gravitational wave driven coalescence of a binary with highly dissimilar component masses is referred to as an extreme mass ratio inspiral (EMRI). It is typically considered "extreme" when the secondary mass is at least a few orders of magnitude smaller than the primary, such that the mass ratio $q\equiv M_2/M_1 \lesssim 10^{-4}$. 
This includes the inspiral of stellar-origin black holes, neutron stars, white dwarfs (and in some cases, stars) into supermassive black holes.
Events in the intermediate mass ratio regime are also possible, such as  mergers of IMBHs with stellar-mass BHs. For the latter, we refer the reader to Chapter 2.  
Compared to more equal-mass binaries, EMRIs are notably different in
their slow and complex orbital dynamics. 
The GW frequency of these events is determined by the mass of the primary BH, making them primary targets for future space-based interferometers. 
EMRIs are particularly exciting given that even a single detection in GWs will provide measurements of SMBH mass and spin to remarkable precision and allow for tests of general relativity in the strong-field regime. 
However, this incredible potential comes with a set of challenges. 
We begin this section with a discussion of why we expect EMRIs to occur, based on several formation channels. 
Within the category of EMRIS, there is a large range in the expected primary mass and and mass ratio due to the varied astrophysical scenarios in which such sources can form. Generally, the primary mass ranges between $\sim 10^5 \, \rm{M}_{\odot}$ to $ \sim 10^8 \, \rm{M}_{\odot} $, while the mass ratio is by definition $q < 10^{-4}$. Nevertheless, it is often useful to refer to a standardised system, the so-called "typical" EMRI, based on the expectation that the centre of our own Milky way is representative of many galaxies in the local Universe. Unless explicitly stated, we assume as customary a primary mass of $10^6\, \rm{M}_{\odot} $ and a $10\, \rm{M}_{\odot}$ secondary whenever we refer to a typical EMRI.\\

In this Chapter, we will approach the topic of EMRIs by starting from their astrophysical origin. In Section \ref{sec:FormationChannels}, we explain several of the expected EMRI formation channels, focusing on some basic physical intuitions and reporting the associated rates. We then shift the focus onto single EMRI systems, take a look at their evolution, GW emission and the possibility of detection in Section \ref{sec:GWdyn}. Finally, we briefly highlight some of the exciting scientific potential of EMRIs in Section \ref{sec:potential}.

\vspace{15mm}

\section{Formation mechanisms and expected rates}
\label{sec:FormationChannels}
\subsection{Dynamical Channels}
\subsubsection{Relaxation in nuclear clusters}
Many super-massive black holes in the local universe are surrounded by nuclear clusters (NCs), dense collections of stars and stellar remnants that are gravitationally bound to regions of order a few pc. Typically NCs contain upwards of millions of solar masses in stars, and a fraction of that in compact objects. The trajectory of an object in a NC is largely determined by the SMBH's gravitational potential, however the large number density facilitates the occurrence of gravitational scattering events, which modify orbital characteristics over long periods. In particular, any given object in a NC will be subject to repeated two-body encounters, which will induce changes to the objects energy and angular momentum in a process known as ``two-body relaxation".
Two body relaxation is thought to be the main formation channel for EMRIs detectable by future mHz detectors. Qualitatively, this scenario works as follows \citep{1997Sigurdsson,2003alex,2005hopman,2013Merritt}: A compact object in a NC is scattered by repeated two-body encounters onto an orbit with low angular momentum, which corresponds to a periapsis close to the central SMBH. As the object swings by the SMBH, it will loose energy through gravitational waves, reducing the orbital semi-major axis and dynamically decoupling from the rest of the NC. The object will then keep loosing energy through GW, circularise and eventually in-spiral into the central BH. 

The main ingredients for a estimate of the event rates for EMRIs are two timescales. The first one is the angular momentum two-body relaxation timescale $T_{\rm{rlx}}$. The value of $T_{\rm{rlx}}$ determines the efficiency with which the cluster can produce high eccentricity EMRI candidates, and depends on the properties of the nuclear cluster, such as the velocity dispersion $\sigma$, the half mass radius $r_{\rm h}$, the steepness $\gamma$ and average number density $n_0$ of compact objects \citep{BinneyTremaine}:
\begin{align}
    T_{\rm{rlx}}&= T_0 \left( \frac{a}{r_{\rm h}}\right)^{\gamma -3/2} \left(1-e^2 \right)\\
    T_0 &= 0.34 \frac{\sigma^3}{\ln \left( \Lambda \right) G^ 2m_{\rm{CO}}^2n_0} \sim \text{few}\, 10^8 \,\text{to}\, 10^9 \, \rm{yr}
\end{align}
where $a$ is the semi-major axis, $e$ the eccentricity and $m_{\rm{CO}}$ the mass of the object, while $\ln (\Lambda)\sim 10$ is the Coulomb logarithm. The second one is the GW-inspiral timescale $T_{\rm{gw}}$, which determines how efficiently GW radiation can dynamically decouple an EMRI candidate from the rest of the cluster. Generally speaking, $T_{\rm{gw}}$ can be approximated by using the seminal result of \citet{1963peters,1964peters}.\footnote{In many astrophysical scenarios however, relativistic corrections to the timescale will lead to more precise results see e.g. \citet{2020zwick,2022acevez}.} Here  we only report the original formula in the limit for very high eccentricities ($e > 0.99$), appropriate for this formation channel:
\begin{align}
    \label{eq:tgw}
    T_{\rm{gw}} &\approx \sqrt{2}\frac{24}{85}\frac{a^4 c^5}{G^3 M m^2_{\rm{CO}}} (1-e^2)^{7/2} 
\end{align}

This timescale is typically evaluated at a periapsis of $4 r_{\rm{S}}$, the parabolic capture orbit (PCO) of a non-spinning BH. Any lower, and the EMRI would likely directly plunge into the SMBH without producing a long lived GW signal. Any higher and GW emission would not significantly affect the  EMRIs semimajor axis in a single passage. 
Setting these two timescales equal to each other yields a length scale, the critical semi-major axis $a_{\rm{crit}}$, which marks the separation between the relaxation- and the GW-dominated regions of the cluster. A candidate with a greater semi-major axis than $a_{\rm{crit}}$ will likely be re-scattered out of its current orbit before it can become a GW source. A candidate with a lower semi-major axis is unlikely in the first place to have its orbit scattered to a low angular momentum. This means that the typical EMRI will have a semimajor axis comparable to $a_{\rm{crit}}$ and a periapsis comparable to the PCO.
Typically, the critical semi-major axis is on the order of 0.1 pc for a Milky Way like NC, although its value depends on many parameters such as the SMBH and compact object masses, cluster mass, density profile and stellar mass function. Crucially, it is several orders of magnitude larger than the typical size of the ISCO around the central black hole, meaning that most EMRI candidates will be produced with very high eccentricities (justifying the use of Eq.~\ref{eq:tgw}).
As the candidate completes many passes around the central BH, gravitational wave emission will tend to circularise its orbit, lowering its semimajor axis and increasing the frequency of periapsis passes until the whole system becomes a mHz source. In many cases, the system will enter the mHz band with a significant eccentricity, typically of the order $\sim 0.5$. Higher eccentricities can be retained for EMRIs orbiting lighter SMBHs, since the mHz band is entered at larger separations (in terms of the Schwarzschild radius), while the opposite is true for heavier systems.
The relaxation formation channel for EMRIs predicts certain event rates, i.e. a certain number of expected EMRIs per galaxy per unit time. These event rates depend on several assumptions and are therefore highly uncertain. Nevertheless, the most recent estimates yield event rates of typically $\sim 10^{-6}$ per galaxy per year. Additionally, the rates can vary by orders of magnitude depending on the SMBH mass, spin and cluster properties. These event rates can be converted into detection rates for LISA by integrating over a cosmological volume and including only EMRIs that fulfill some SNR requirement. Expected detection rates range anywhere from none to thousands of EMRIs throughout LISA's observation run, with most models predicting a median expectation of tens to hundreds of detectable events \citep{2004gair,2007lisarev,2010preto,2011pau,2017lisarev}.

\subsubsection{Binary disruption (Hills mechanism) or binary capture}
The above estimates typically assume that NCs only consist of single stellar systems. If binary or multi-body stellar systems are present in an NC, EMRIs can also form via binary disruption, a process more formally referred to as the Hills mechanism, crediting the original work by \citet{1988Natur.331..687H,1991AJ....102..704H}. 
This occurs when a binary travels deep enough in the center of the nucleus for the binary to become tidally separated by the gravitational force from the SMBH, leading to one component becoming bound to the SMBH and the companion being ejected from the system. This mechanism is invoked to explain hyper-velocity stars traveling through the halo of the Milky Way \citep{2015ARA&A..53...15B}. 
In some sense it is `easier' to form EMRIs via this channel, because 
the cross section of binary capture is larger than for the single BH case.  
Consider a binary of total mass $m_{\rm bin}$ at separation $a$, which will be disrupted by the central SMBH of mass $M_{\rm BH}$ if it passes within the binary tidal radius:
\begin{equation}
    r_{\rm bin} \sim \left(\frac{M_{\rm BH}}{m_{\rm bin}} \right)^{1/3} a,
\end{equation} 
where we neglect a numerical factor in the cube root that depends on whether the binary is prograde or retrograde \citep{1991Icar...92..118H}.
A typical binary is disrupted if it enters within $\sim10$s of AU from an SMBH, compared to the $\sim0.1$ AU required for two-body relaxation.
Additionally, the resulting bound BH is found on a lower eccentricity, smaller pericenter orbit, such that the nascent EMRI is less vulnerable to subsequent perturbations from surrounding stars and thus more guaranteed to eventually coalesce with the SMBH. 
Even if the binary fraction of the NC is only a few percent, this channel is expected to produce EMRIs at a rate comparable to or greater than single stellar scattering \citep{2005ApJ...631L.117M}. 

A binary can avoid disruption close to the SMBH if it is sufficiently tightly bound. In this case, it may survive coalescence to form a \emph{binary}-EMRI, or a hierarchical triple system which produces a unique multi-frequency signal if detectable by space-based interferometers \citep{2018CmPhy...1...53C}. For a central SMBH of $10^6 M_{\odot}$, this requires that the stellar binary separation be less than $a\lesssim 5\times10^{-3} AU$ (or $5\times10^4 $ gravitational radii, for a binary of $10 M_{\odot}$ BHs) to survive a large portion of the inspiral. This rate again depends on the binary fraction of the NC, but optimistic estimates suggest it can be as high as 10 percent of the total EMRI population.

\subsubsection{Second Massive Black Hole}
NCs should not necessarily be considered isolated objects. Over the relevant timescales of billions of years, their hosts participate in the complex dynamics of cosmic structure formation, in which galaxies repeatedly merge to form larger ones. Whenever two galaxies merge, their SMBHs slowly start sinking towards the centre, eventually coming to small enough separations to be a able to influence each other gravitationally. For simplicity, imagine a secondary SMBH approaching a primary SMBH with an intact nuclear cluster. In this scenario, the orbit of individual compact objects within a nuclear cluster will be perturbed by the approach of the former.
There are two principal ways by which the secondary SMBH can produce additional EMRI candidates (around the primary SMBH). Firstly, orbital dynamics can become chaotic whenever its gravitational influence is strong enough to compete with the primary SMBH, naturally forcing some compact objects to high eccentricities. Secondly, secular resonant behaviours in hierarchical triples can induce so called Kozai-Lidov oscillations \citep{1910vonzeip,1962lidov,1962kozai}, in which orbital inclination is exchanged for orbital eccentricity, also forcing some objects onto low periapsis orbits. Both of these effects will produce additional EMRI candidates, which must then still fulfill the criteria determined by the standard two-body relaxation process and the loss of energy through GWs \citep{2014bode,2022naoz}.

In general, EMRI formation channels that involve the presence of a secondary SMBH are extremely efficient, producing sharp spikes in the estimated event rates. However, the increased efficiency only lasts for a short time, as the secondary SMBH is likely to further sink towards the centre, tidally strip the nuclear cluster and eventually disrupt it completely \citep{2022mazzolari}. Thus, the actual rate of these formation channels must be averaged over the long stretches of time between galaxy mergers, in which an appropriate configuration of bodies is missing. Nevertheless, recent estimates concur that EMRIs induced by the presence of a second SMBH might still be comparable to the standard dynamical channel, producing several dozens of detectable events per year.

\subsubsection{Supernova kicks}
Stars in NCs evolve along the main sequence and can end their lives by undergoing a supernova. During the violent collapse, small asymmetries in the infalling material can induce so-called natal kicks, i.e. impart a peculiar velocity to the supernova remnant, as a consequence of momentum conservation. Interestingly, the typical kick velocity $v_{\rm k}$ of main sequence stars is of the order of several 100 km/s \citep{1970Sk,1983anderson,2005Hobbs}, comparable to the velocity dispersion of a Milky way-like nuclear cluster. Natal kicks are therefore likely to significantly perturb the trajectory of a newly birthed compact object, and can occasionally scatter it into a low angular momentum orbit producing an EMRI candidate \citep{2019Elisa}.

Once such a candidate is produced, it must still fulfill the criteria for dynamical EMRIs, i.e. its orbit must not be efficiently affected by two-body relaxation, nor come so close to the central SMBH that a direct plunge occurs. Only a very small fraction of supernova remnants are able to successfully produce an EMRI, approximately $10^{-4}$ to $10^{-7}$ depending on their stellar component of origin. This leads to an estimated SN-EMRI rate of $\sim10^{-8} \rm yr^{-1}$ per Milky Way-like galaxy.
Nevertheless, supernova origin EMRIs could comprise a fraction of the total detection rate, simply due to the large number of stellar explosions over the relevant timescales of billions of years.

\subsection{EMRI generation in active galactic nuclei}
\label{sec:wetemris}
In addition to interacting with occasional stellar-origin objects, SMBHs are also known to accrete gas episodically throughout their lifetimes. Gas accretion is a critical process for the growth of SMBHs, particularly to explain those that reach very high masses. The process can release a massive amount of energy, thus powering emission that we observe as active galactic nuclei (AGN)\footnote{
AGN exhibit a large diversity of emission features across the EM spectrum. Not all components and the observed diversity are fully understood, but the accretion of gas via a rotationally supported configuration is a widely accepted component. We refer the reader to \citet{1992apa..book.....F} to learn more. 
}. 
This includes the most electromagnetically luminous objects in the Universe, quasars\footnote{Quasars are a subclass of AGN, referring to the most distant, luminous sources where we cannot resolve the galactic environment.}, which are currently observed up to redshifts as high as $z\sim7$ \citep{2021ApJ...907L...1W}. 
Currently we can observe that a fraction ($\sim 10\%$) of galactic nuclei are gas-rich.
 Thus, from an EMRI standpoint, assuming the stellar dynamics discussed above is unaffected by the presence of gas, then we should expect at least the same fraction of EMRIs to occur in AGN. However, it is also likely that the presence of gas in the nucleus affects stellar dynamics, leading to a potential boost in the EMRI rate for sufficiently gas-rich nuclei.

In most cases, the accretion flow on sub-parsec scales in an AGN is electromagnetically spatially unresolvable\footnote{
The exceptions are observations that utilize reverberation mapping in the X-ray band, which can provide properties of the inner accretion flow. 
Additionally, sub-mm interferometry
has successfully spatially resolved the small scale accretion flow around the SMBH in our own Galactic Center and in M87 \citep{2019ApJ...875L...1E,2022ApJ...930L..12E}.   
However these SMBHs have a low accretion rates (i.e. the luminosity is significantly lower than the Eddington luminosity $L \sim 10^{-5} - 10^{-8} L_{\rm Edd}$) and thus these measurements probe very different types of low density, radiatively \emph{in}efficient accretion flows.}, and thus we rely on models to predict properties of the gas and its morphology.
The most common models are categorized by the rate of mass inflow. 
A simplified accretion model can conveniently be described by analytical solutions following certain assumptions:
(i) the disk is in steady-state, described by a constant accretion rate $\dot{M}$,
(ii) the mass of the disk is small compared to the central object, and thus the angular velocity is Keplerian, 
(iii) the rate of gas inflow is governed by viscous transport of angular momentum, which is assumed to be driven by turbulence or magnetically driven instabilities and is  parameterized by a parameter $\alpha$, 
and (iv) the disk is able to efficiently radiatively cool the energy from internal heating generated during mass transport. 
From mass and energy conservation, 
we can then describe the surface density of an accretion disk at separations within the critical radius for EMRI capture by
\begin{equation}
    \Sigma(r) \approx 7300 
    \left( \frac{0.01}{\alpha} \right)
    \left( \frac{f_{\rm Edd}}{0.1} \right)
    \left( \frac{M_{\rm BH}}{10^6 M_{\odot}} \right)^{1/2}
    \left( \frac{10^{-2}}{h/r} \right)^2
    \left( \frac{0.01 {\rm pc}}{r} \right)^{1/2} {\rm g~cm^{-2}} .
\end{equation} 
Here we have further simplified the disk model for a constant aspect ratio $h/r=0.01$ and a constant accretion rate that is a fraction $f_{\rm Edd}$ of the Eddington rate (with a radiative efficiency set to 10 percent). Note that to convert this to a volumetric density, we will need to drop the former assumption and solve for $h$ self-consistently with knowledge of the gas cooling rate to obtain the disk's vertical structure.
The takeaway message is that for luminous AGN, the accretion disk can be relatively dense and highly azimuthally supersonic ($v_{\phi}\gg c_s$) in regions of interest where stars/BHs will interact with the disk. 
A canonical work by \citet{1973A&A....24..337S} expands on this model to derive analytical solutions for accretion flow that have been widely adopted for many works.   
The model serves as a convenient tool, but unfortunately it is too simple to explain the variability and diversity of emission features in observed AGN, which consist of additional features beyond the accretion disk that we do not describe here.  
However, it nevertheless provides insight on how the presence of gas in such a configuration can influence the rate of stellar interaction with the SMBH.  \\

\textbf{Orbit capture of nuclear cluster stars/BHs}:
A star in the nucleus of an AGN that intersects the accretion disc throughout its orbit will lose energy and momentum with each passage due to drag forces, leading to a change in inclination and eccentricity. 
Provided enough intersections, even initially high eccentricity orbits can become near circular and co-rotating within plausible AGN lifetimes of $\sim 1-10^2$ Myr \citep{1991MNRAS.250..505S, 1983ApJ...266..502N,1993ApJ...409..592A}. 
Thus the presence of a disk can significantly alter the distributions of eccentric orbits in the NC, as long as the AGN disk is sufficiently dense. This is a more important effect for stars that have a larger cross section and are hence more susceptible to drag forces, unlike compact objects \citep{1995MNRAS.275..628R,2020ApJ...889...94M}.

The timescale for a star to become aligned with the disk is approximately the time it takes for a star to cumulatively interact with disc gas equal to its own mass. 
Within one orbit crossing, the intersected disk mass is approximately
\begin{equation}
    \Delta m_{\rm gas} = \pi R_*^2 \Sigma,
\end{equation}
for a star of radius $R_*$, where $\Sigma$ is the disk surface density. For a compact object, the radius term reduces accordingly (gravitational drag (dynamical friction) becomes important rather than hydrodynamical (surface area) drag). One can combine this expression with a disk model
to obtain an order of magnitude estimate of the required number of intersections needed for disk capture for a given star. 
From this we can see that the structure of the star or compact object plays a role in determining how efficiently a stellar orbit is trapped within the disk, in addition to the disk properties. 
Even with the simplified thin-disk model, it is clear that in at least some regimes the presence of an accretion disk can affect the stellar dynamics, either for stars that are scattered in via two-body encounters, and additionally for those that intersect the disk at farther distances, beyond which they would not have become EMRIs via the dynamical formation channels.
In general this process will increase the number of stars that become strongly bound to the SMBH, especially for stars that plunge through the disk on highly eccentric orbits, during which they interact with a denser region of the disk at periapse. This interaction can dramatically decrease the eccentricity of the orbit, which in turn leads to a decrease in inclination \citep{1995MNRAS.275..628R}. \\

\textbf{In-situ formation of stars/BHs}:
 Accretion discs can become subject to gravitational instability if the self-gravity of the gas overcomes the pressure and rotational support \citep{1964ApJ...139.1217T}. 
 This typically occurs if disks are sufficiently massive and/or sufficiently thin, when they extend beyond $\sim 0.01$ parsecs. 
In these regions the instability generates spiral density waves which alter the disk structure and evolution  \citep{1987Natur.329..810S,1989ApJ...341..685S,2003MNRAS.339..937G}.
 Under the right conditions gravitationally unstable discs will cool and fragment, leading to fragmentation and star formation. 
 Initial conditions of the in-situ stellar population are dependent on the gas properties, and stars formed in the hotter, denser environments of AGN (as opposed to the molecular clouds) are expected to follow a top-heavy mass distribution. 
 Further evidence of this process lies in the stellar discs observed around Sgr A*, which can be explained by gravitationally-induced fragmentation of an infalling cloud \citep{2003ApJ...590L..33L}. 
 For AGN, which are expected to contain a higher amount of gas to explain the bright emission, fragmentation can result in star formation in an environment that remains gas-rich. 
In this case, subsequent evolution of stars in the remaining accretion disk can lead to 
metal enrichment of the gas and embedded compact remnants, many of which can eventually coalesce with the central SMBH as the disk is accreted. This channel can lead to a substantial boost in the EMRI rate compared to a dry galactic nucleus, reaching as high as $10^{-4} ~\rm yr^{-1}$ events per AGN--a rate comparable to or greater than that driven by two-body relaxation \citep{2023MNRAS.tmp..731D}. Determining this rate depends on the population of AGN within the EMRI detection volume as well as details regarding stellar migration during the disk evolution, both of which lead to a wide range of uncertainty. \\

\textbf{Once embedded and co-rotating with the disk}, a star will \emph{migrate} or exchange angular momentum, expanding or shrinking its orbit. 
This is due to excitation of density waves 
which back-react on the embedded star's orbit\footnote{Much of the seminal work on migration outcomes has been done in the context of planet migration in protostellar disks. For further reading on the derivation of migration torques, we additionally refer the reader to \citet{2020apfs.book.....A}.} \citep{1979ApJ...233..857G,1978ApJ...222..850G,1980ApJ...241..425G,1989ApJ...347..490W}. 
The theory suggests that embedded stars/BHs should rapidly migrate inward, providing a simple pathway for inevitable coalescence with the central SMBH. 
The picture becomes more complex when one considers deviations from the linear theory which complicate migration. These include stochastic torques due to disk turbulence, the effect of feedback from stars or accreting BHs, complexities in accretion disk structure,  mutual interactions between embedded objects, or the interplay of all of these effects simultaneously. 
One can also turn this challenge into an intriguing possibility:  if detected, the orbital characteristics of even a single gas-embedded EMRI (or the distribution over a population) will provide insight into these poorly constrained environments. \\

\begin{figure}
\centering
\includegraphics[scale=0.45]{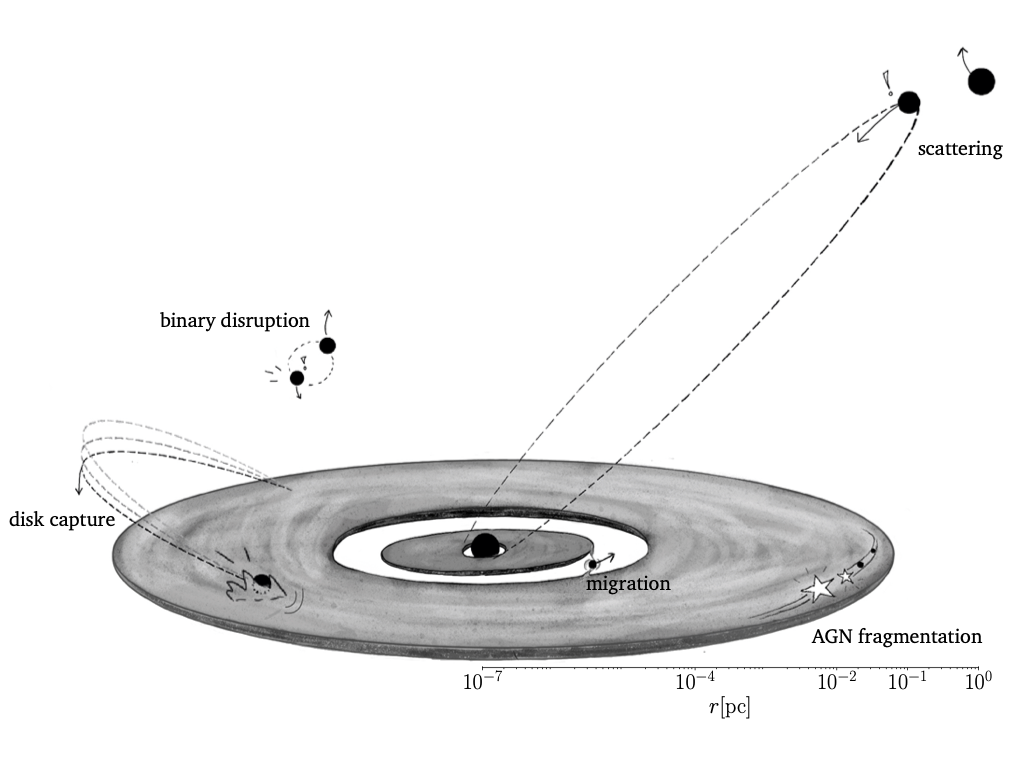}
\caption{Illustration of the various EMRI formation channels in a galactic nucleus and the relevant physical scales.  }
\label{fig:schematic}
\end{figure}

To summarize, the presence of accretion disks is expected to increase the interaction rate of stars/BHs in the nucleus, either by grinding down their orbits or forming stars in-situ. While AGN only comprise a fraction of galactic nuclei, 
the possibility exists for 
gas-embedded EMRIs to potentially be more common than their dry counterparts. 
Ultimately this will depend on the prevalence of active SMBHs in the low redshift Universe (within the detection volume,  $z\sim1-3$), and details of the processes underlying EMRI generation which need to be constrained.

\subsection{Additional types of EMRIs}

\subsubsection{IMRIs due to sBH mergers in disks}
Accretion disks can also facilitate binary interactions between embedded stars and black holes. 
This is a possible formation channel for high frequency BH mergers that are detectable by ground-based detectors \citep{1993ApJ...408..496M,2012MNRAS.425..460M}. Since we focus on inspirals with the central SMBH in this part of the chapter, we only note here that such hierarchical mergers may lead to intermediate mass BH (IMBHs) embedded in accretion disks, which can potentially coalesce with the central SMBH producing even louder milliHz GW events.

\subsubsection{Inspirals of stars or sub-stellar objects (XMRIs)}
The discussion of EMRIs typically focuses on the inspiral of a stellar-origin BH, given that these sources are more likely to be detected than EMRIs composed of lighter objects, considering their louder GW strain. In principle, however, stars and compact remnants of various types can be scattered toward a SMBH. Whether or not they survive to inspiral due to GW emission depends on their ability to survive the SMBH's tidal forces. These forces can be strong enough to overcome the binding energy of a star, thus disrupting the star before it becomes a substantial GW source. 

Consider a star of mass $M_*$ and radius $R_*$, which follows a mass-radius relation $R_{*}\propto M_*^{\alpha}$. 
The tidal disruption radius near an SMBH of mass $M_{\rm BH}$ is \citep{1975Natur.254..295H}
\begin{equation}
r_{\rm tidal} \approx R_* \left(\frac{M_{\rm BH}}{M_*} \right)^{1/3}.    
\end{equation} 
Whether a star survives disruption when encountering the SMBH will depend on its internal structure and the SMBH mass. 
Specifically, a star will avoid disruption and coalesce into the SMBH if the tidal radius is within the event horizon or the plunge radius, i.e $r_t < 6 G M_{\rm BH}/c^2$ for a Schwarzschild BH. 
The above expression reveals that for more massive SMBHs (e.g. $M_{\rm SMBH}\gtrsim {\rm few}\times 10^7 M_{\odot}$), stars within an increasingly large mass range will coalesce before disrupting. However, in these cases the resulting GW signal is produced at lower frequencies outside of the realm of mHz detectors.  
For lower mass SMBHs such as that in our Galactic Center, inspirals of compact and substellar objects are possible. 
Brown dwarfs are supported by electron-degeneracy pressure, and thus can be relatively compact with radii $R_{\rm bd}\propto M_{\rm bd}^{-1/3}$.  
 This implies that they will survive disruption around Sgr A*, making it possible to produce  EMRIs with eXtremely small mass ratios (XMRIs) which will be detectable by mHz detectors. 
A consequence of the extremely low mass is that the inspiral produces GWs with a very weak strain, such that these events can only be detected in the Galactic center of the Milky Way or nearby galaxies. Depending on the number density of brown dwarfs in our own Galactic center, there may be as many as $\sim 10$ XMRIs present at any given time \citep{2019PhRvD..99l3025A,2019A&A...627A..92G}). 

There also exists a mass range of stars that will produce GWs in a detectable frequency band for at least some time before being disrupted. For a better understanding of electromagnetic and gravitational wave emission from tidal disruption events, we refer the reader to Part III of this chapter.

\subsection{Peculiarities, similarities, and differences of dry and wet EMRIs}

The variety of formation channels for EMRIs leads to a diverse range of orbital properties once EMRIs become GW-dominated and detectable by space-based interferometers.  
For a visualization of the primary formation channels, we point the reader to Figure~\ref{fig:schematic}.
EMRIs from gas-assisted formation channels are typically predicted to have low eccentricity compared to those which form via two-body relaxation. 
The consequence of this is that gas-embedded sources 
evolve on much simpler and long-lived trajectories. Perhaps this leads to higher chance of detection, as we will discuss in Section~\ref{sec:GWdyn}, while sacrificing some of the complexity which allows for strong field tests of GR. 
Note that we expect to constrain EMRI eccentricity and inclination with high precision (assuming we address data analysis challenges), which will make it possible to distinguish even low-eccentricity events (e.g. $e \lesssim 10^{-3}$ or lower, when measured at plunge, see \citet{2017lisarev}) from zero-eccentricity events.

The orbital inclination of EMRIs with respect to direction of SMBH spin will also differ between certain formation channels. 
EMRIs that form via accretion-driven channels should nearly equatorial, if the SMBH spin is driven towards alignment with the angular momentum of the accretion disk. (The latter requirement is generally expected but not always guaranteed, depending on the disk orientation and lifetime \citep{2005MNRAS.363...49K,2008MNRAS.385.1621K}.) 
EMRIs that form via two-body relaxation or binary capture, on the other hand, should show no correlation in inclination with respect to the SMBH spin. 
Thus while both accretion-driven channels and binary disruption can produce events with low eccentricity, inclination measurements can help to distinguish between these channels, providing insight into the formation mechanisms and properties of the host galaxy. 
Constraining the formation channel with certainty for a single event may be challenging, but with multiple events, the distribution of these parameters will be meaningful for understanding dynamics in galactic nuclei.

\subsection{Multimessenger prospects of accretion-driven EMRIs}
The gaseous environment of EMRIs in AGN raises the possibility of \emph{multimessenger} detections, if characteristic electromagnetic emission is generated prior to, during, or after the GW event is detected. For `typical'\footnote{The expected mass range of a typical EMRI in an AGN disc remains uncertain, but recent work suggests that wet EMRIs may not be driven to significantly higher masses, given that 
BH accretion is likely limited by mechanical feedback \citep{2022ApJ...927...41T}. } EMRIs, it is unlikely that the BH will perturb an accretion disk significantly enough to produce emission that is distinguishable from ubiquitous AGN variability \citep{1997ARA&A..35..445U}. However, stellar EMRIs may produce unique periodic emission when mass is stripped from the companion (e.g. \citealt{2022ApJ...926..101M,2023MNRAS.524.6247L}). Quasi-periodic eruptions (QPEs) are observed in multiple low mass galactic nuclei \citep{2023MNRAS.526L..31K}. While their origin is still unknown, mass stripping from a stellar or white dwarf companion is a viable candidate \citep{2021Natur.592..704A}.
A similar albeit non-periodic multi-messenger signal will exist for tidal disruptions that occur within detector bands, which should produce a characteristic flare (for a thorough discussion, we refer the reader to Section III of this Chapter).
For IMRIs, the IMBH exerts a stronger gravitational influence on the surrounding gas, and the formation of a gap in the disk can lead to a decline in the AGN luminosity coincident with the final coalescence \citep{2013MNRAS.432.1468M}. 

Besides detecting EM counterparts for a single event, it is also possible to use the characteristics of an EMRI to narrow down the list of possible host galaxies in the localization volume. 
Perhaps with multiple events, AGN will be statistically over- (or under-) represented in the collective localization volume, as has been shown for LIGO/Virgo detections \citep{2017NatCo...8..831B}, which will hint towards the dominant mechanism for EMRI formation.

\section{Basics of EMRI detection}
\label{sec:GWdyn}

\subsection{EMRI Dynamics and Waveforms}
The dynamics of EMRIs are a truly fascinating and complex field, since GR introduces several qualitative differences with respect to simple Newtonian orbits. Crucially, the instantaneous trajectory of an EMRI completely determines its emission of GWs, which will therefore encode information about metric along the EMRIs path. The extreme mass ratios that characterise such systems mean that EMRIs can persist for thousands of orbits in the close vicinity of the primary's Schwarzschild radius, in a regime in which strong curvature effects become relevant. In particular, EMRI trajectories are strongly affected by both the Schwarzschild and Lense-Thirring precessions even on short (orbital) timescales. The former manifests itself as a periastron advance, and is primarily caused by $\sim 1/r^2$ corrections to the potential of a Schwarzschild BH with respect to a Newtonian point mass. The latter can be thought of as a nodal precession, i.e. a precession of the orbital plane, primarily caused by the induced quadrupole gravitational potential of the central spinning SMBH. To get a feel for the size of the two aforementioned precession rates, we report their value in degrees per completed orbit \citep{1973gravity}:
\begin{align}
\Delta \omega &\sim \frac{6 \pi G M}{ a(1-e^2) c^2} \approx 54^{\circ} \times \frac{10 r_{\rm S}}{p} \\
\Delta \Omega &\sim  4 \pi \xi\left(\frac{G M}{a(1-e^2)c^2}\right)^{3/2} \approx  8^{\circ} \times \xi \left(\frac{10 r_{\rm S}}{p}\right)^{3/2},
\end{align}
\noindent where $\xi$ is the dimensionless spin parameter and $p=a(1-e^2)$ is the semi-latus rectum.
For the typical EMRI with moderate eccentricities and a peri-apis of $\sim 6 r_{\rm{S}}$, both Schwarzschild and Lense-Thirring precessions are of the order of tens of degrees per orbit, clearly manifesting the strong difference of relativistic orbits from simple Keplerian ellipses. Furthermore, the system's emission of GWs is associated with energy and angular momentum fluxes. These cause a back reaction onto the trajectory that is simply not present in Newtonian mechanics, adding dissipative terms which tend to circularise and shrink the orbit, further modifying the dynamics and ultimately leading to the merger.

\begin{figure}
\centering

\begin{tabular}{c|c|c}
  Order   & Qualitatively new features & Further reading \\ 
  \hline \\
     $\left(\frac{v}{c} \right)^{0}$  & Newtonian orbit & -- \\
     $\left(\frac{v}{c} \right)^{2}$   & Periastron precession & \citet{1916Einstein}\citet{1973gravity}\\
     $\left(\frac{v}{c} \right)^{3}$   & Spin-orbit coupling/Lense-Thirring precession & \citet{1918Lense}\citet{1960schiff}\\
     $\left(\frac{v}{c} \right)^{4}$  & Spin-spin coupling/spin precession & \citet{2014blanchet}\\
     $\left(\frac{v}{c} \right)^{5}$  &Energy and momentum fluxes, GWs & \citet{1963peters} \citet{1964peters}\\ 
     $\left(\frac{v}{c} \right)^{6}$  & Higher GW modes & \citet{1980thorne}\\
     $\left(\frac{v}{c} \right)^{8}$ & Post Newtonian tails & \citet{2014blanchet}\\
     ...  &...&
\end{tabular}
\includegraphics[scale=0.605]{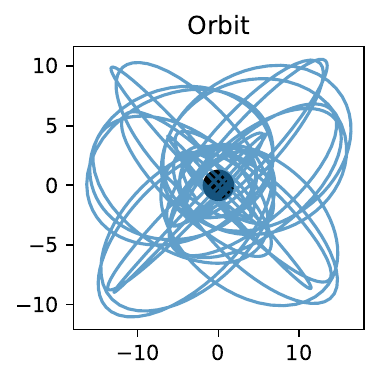} \includegraphics[scale=0.605]{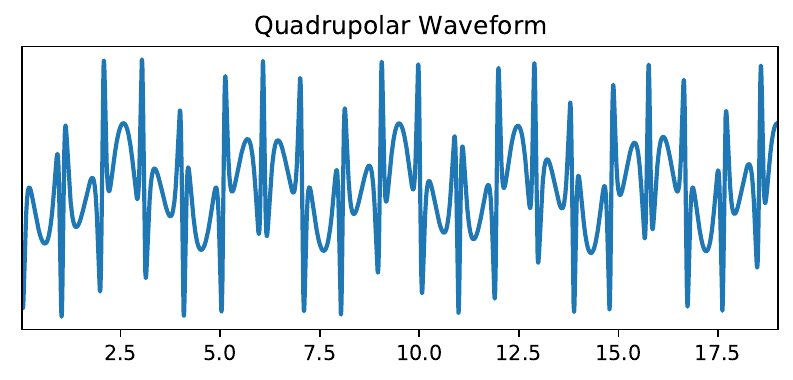}
\caption{By approximating GR to higher and higher precision, more and more complexity is added to the starting point of a simple Newtonian orbit and a sinusoidal GW. Here we show the trajectory of a test mass orbiting a spinning primary BH on a eccentric and inclined orbit for twenty periods. We also show the resulting quadrupolar part of the GW signal, projected along some arbitrary axis. The GW clearly shows high amplitude peaks associated with pariapsis passages, as well as a slow modulation caused by the precession of the orbital plane.}
\label{fig:15}
\end{figure}

In order to be confidently detected and for their parameters to be reconstructed without bias, EMRIs will require extremely precise models of their dynamics and their GW emission.
In order to understand this statement, we introduce a very useful quantity used both in data analysis and GW astronomy: the total \textit{accumulated phase} of a waveform. Consider the the total phase of a typical circular EMRI orbiting in the mHz band over the course of an observation window $T$:
\begin{align}
    \phi_{\rm{tot}} = \int_0^{T} f dt \approx f_{\rm{mc}}T + \frac{1}{2} \dot{f}T^2 + \mathcal{O}[T^3]
\end{align}
Where we expanded in the observation time $T$, i.e we assumed that the frequency $f$ of an EMRI GW only evolves slowly. As expected, the total accumulated phase is mostly determined by the frequency at the beginning of the observation. However, an additional term proportional to the rate of change of the frequency $\dot{f}$ appears as the leading order correction in $T^2$. Crucially, $\dot{f}$ corresponds to the evolution of the EMRI's orbital parameters caused by energy and angular momentum fluxes. Here we report the orbit-averaged expression for $\dot{f}$, expanded at lowest order in the the small quantity $v/c$ \citep{1963peters}:
\begin{align}
    \dot{f} = q f^{11/3} \frac{96}{5}\frac{G^{5/3} M^{5/3}}{c^5} \left(1 + \mathcal{O}\left[\left(\frac{v}{c} \right)^7 \right]\right) \sim 3\times 10^{-14}\, \left[\rm{s}^{-2}\right]
\end{align}
where $v$ is the typical orbital velocity of the system, we simplify for small mass ratios $q << 1$ and also neglect to specify the red-shift dependence of the formula. Evaluating the total accumulated phase for a typical EMRI, we find:
\begin{align}
    \phi_{\rm{tot}} \sim 1.2 \times 10^5 + 216.1\left(1 + \mathcal{O}\left[\left(\frac{v}{c} \right)^7 \right]\right)
\end{align}
This expression qualitatively tells us that, even though EMRIs do not evolve rapidly over the course of an observation time, GW emission is still a significant contributor to the total phase of the event. Moreover, it tells us that very high order energy radiation models are required to determine the phase to a small fraction of $2\pi$, which is the typical accuracy threshold for sources with an SNR of $\sim 10$ to $\sim 100$ \citep{2004barack,2015chua}. In this particular example, the secondary compact object is orbiting with a speed of approximately 0.3$c$, which implies that terms of the order $\sim (v/c)^{12}$ are still expected to produce a clearly significant contribution to the total phase of the signal! Similar precision is also required when modelling the conservative evolution (i.e. without considering GW induced fluxes) of the system, since effects such as periastron or nodal precession may also contribute to the total phase accumulation depending on the orientation of the system.

\subsection{Approximation schemes}
The arguments above show that models of EMRI dynamics and waveforms must achieve an incredible accuracy to hope to precisely match real signals. The field of numerical relativity, i.e. the direct integration of the Einstein field equations could in principle provide such accuracy. However, there are many characteristics of EMRIs that do not lend themselves to be treated numerically in the framework of GR. Firstly, while EMRIs are expected to complete several tens of thousands of cycles in the mHz band, their phase must be tracked with sub-radian precision to maximise the SNR of the event. This imposes extremely tight constraints on the numerical accuracy of the integration. Secondly, the mismatch in mass between the primary SMBH and the secondary compact object introduces a large difference in the required discretisation scales. In principle, the grid size around the secondary must be chosen such that it is at least comparable to its gravitational radius, $10^{-4}$ to $10^{-7}$ times smaller than the optimal grid size to describe the primary SMBH \citep{2005pretorious,2018maggiore}.

While both of these effects drastically increase the computational time required to produce a single accurate waveform, the real killer for the prospects of numerical relativity is simply the vastness of parameter space that has to be covered in order to exhaust all possible EMRI configurations. Both component masses can individually vary by several orders of magnitude, ranging from a few $10^4$ M$_{\odot}$ to several $10^7$ M$_{\odot}$ for the primary and from a few M$_{\odot}$ to several $10^2$ M$_{\odot}$ for the secondary. Furthermore, the system's Keplerian orbital elements are essentially unconstrained, since EMRIs can enter the mHz band with a wide range of semi-major axes, eccentricities and inclinations. Additionally, the relative orientation of the primary's spin to the orbital angular momentum is also likely to be random (especially in the dynamical formation channels). Finally, the orientation of the source and the observer's line of sight requires an additional three free parameters to properly project the GW signal onto the detector frame.

Thus, the problem of EMRI waveforms clearly requires a different approach, namely the use of analytical approximations to the fully relativistic orbital dynamics. Analytical approaches have the obvious advantage of being much more efficient than direct numerical integration. However, their precision is ultimately limited by the mathematical challenge of treating the field equations of GR. Here we briefly present a simplified version of the standard procedure to generate approximate waveform templates, known collectively as Kludge\footnote{The literal meaning of the word ``kludge" is quite appropriate to describe this particular technique.} waveforms or simply post-Newtonian waveforms \citep{2004barack,2007babak,2011sopuerta,2018maggiore}. Note however that many additional methods are being explored in the literature.

\subsubsection{Constructing an approximate waveform}
The most straightforward scheme to produce an approximation to an EMRI waveform is to start with a simple Newtonian orbit and perturbatively add more and more analytical relativistic effects. To illustrate the procedure, we start with a Newtonian solution to the trajectory of an EMRI, $\vec{r}_{\rm N}$, characterised by a semi-major axis $a$, an eccentricity $e$, a true anomaly $\phi$ and some additional orbital elements $\vec{\theta}$. The lowest order GW emission $h_{i j}$ of such a trajectory can be approximated by the famous quadrupole formula:
\begin{align}
    h_{ij}^{\rm{Q}}\left[ \Vec{r} \right] = \frac{G}{c^4 D}\frac{d^2}{dt^2}\mathcal{I}_{ij}
\end{align}
where $D$ is the luminosity distance of the source, $\mathcal{I}_{ij}$ is the mass quadrupole tensor of the EMRI system associated to the trajectory $\Vec{r}$, and we neglect to specify any particular gauge choice. However, the generation of GWs is associated with an energy and angular momentum flux, which must affect its semi-major axis and eccentricity as a consequence of energy balance. With some clever calculation \citep{1964peters}, one can can derive a set of evolution equations for $a$ and $e$ from the quadrupole formula, which are strongly proportional to the separation of the binary itself:
\begin{align}
    \frac{da}{dt} &\propto \frac{G^3}{c^5}\frac{1}{a^3}\\
    \frac{de}{dt} &\propto \frac{G^3}{c^5}\frac{1}{a^4}
\end{align}
By combining the trajectory $\vec{r}_{\rm{N}}$, the evolution equations $\dot{a}$ and $\dot{e}$ and the quadrupole formula, one can produce a first consistent approximation to the expected GW of a binary source, often called the "Newtonian waveform", in which the dynamics are accurate up to order $(v/c)^0$ while GW radiation is accurate to $(v/c)^5$ order.
The next upgrade from Newtonian waveforms is achieved by adding the first conservative relativistic modification to the trajectory of the EMRI:
\begin{align}
    \vec{r} \to \vec{r}_{\rm N} + \frac{1}{c^2}\vec{\rho}
\end{align}
which was originally found by Einstein and collaborators \citep{1915einstein,1916Einstein} and essentially introduces the notorious periastron advance also seen in the motion of Mercury. The resulting GW is calculated according to:
\begin{align}
    h_{ij} \to h_{ij}^{\rm{Q}}\left[\vec{r}_{\rm N} + \frac{1}{c^2}\vec{\rho} \right] + h_{ij}^{\rm{C8}}\left[\Vec{r}_{\rm{N}}\right]
\end{align}
where $h_{ij}^{\rm{C8}}$ are the GW contributions associated to a mass octupole and a current quadrupole, which are suppressed by an additional factor $(v/c)$ with respect to the leading quadrupole radiation. Note how the latter must only be evaluated for the Newtonian part of the trajectory, since any $h_{ij}^{\rm{C8}}$ contribution arising from the trajectory perturbation $c^{-2}\vec{\rho}$ would then be suppressed by a factor $c^{-3}$. Finally, the evolution equations for $a$ and $e$ can be updated to include the additional fluxes associated to $h_{ij}^{\rm{C8}}$ and to $h_{ij}^{\rm{N}}[c^{-2}\rho]$. Collecting all the new terms, we now have a set of equations that consistently describe the evolution of the EMRI and its GW radiation by including relativistic effects up to modifications of order $(v/c)^2$ with respect to a Newtonian waveform. Schematically, we can summarise the results as folows:
\begin{align}
  \text{Trajectory:}&  &\Vec{r} = \vec{r}_{\rm N} + \frac{1}{c^2}\vec{\rho}\\
  \text{Periastron advance:}&  &\frac{d}{dt}\Vec{\theta} = \left(\frac{d}{dt}\Vec{\theta}\right)_{1/c^2}\\
    \text{Energy flux:}& &\frac{da}{dt} = \left(\frac{da}{dt}\right)_{\rm{N}} + \left(\frac{da}{dt}\right)_{1/c^2}\\
     \text{Ang. momentum flux:}& &\frac{de}{dt} = \left(\frac{de}{dt}\right)_{\rm{N}} + \left(\frac{de}{dt}\right)_{1/c^2}\\\nonumber \\
     \implies \text{Waveform:}& &h_{ij} = \left(h_{ij}\right)_{\rm{N}}+ \left(h_{ij}\right)_{1/c} + \left(h_{ij}\right)_{1/c^2}
\end{align}
where we have always explicitly reported the additional $(v/c)$ factors relative to the leading order expression in the subscripts. The result of this scheme is a more accurate waveform than what was achieved previously, in which the dynamics are accurate to $(v/c)^2$ order and the GW radiation is accurate to $(v/c)^{7}$ order. It is thus called a first order "post-Newtonian" (PN) waveform.

In principle, this scheme could be repeated over and over, adding new relativistic effects until satisfactory precision is achieved. However, in reality one quickly reaches a point where an analytical treatment of the orbital trajectory becomes exceedingly complex, and further simplifications are required. Particular care is required when improving the waveform from the first PN order to the subsequent 1.5 PN, i.e. dynamics to $(v/c)^3$ and radiation to $(v/c)^8$. The 1.5 PN order not only introduces dynamics related to BH spin, but so called radiative "tail" terms become important. The latter correspond to non-linear corrections of the metric itself, i.e. of order $h_{ij}^2$, that must be added in a consistent perturbative treatment of the field equations \citep{2014blanchet,2018maggiore}. They are notoriously hard to compute analytically, and indeed closed form solutions do not exist for arbitrary eccentricities. Thus, state of the art waveform models are often required to augment the analytical results with interpolations from numerical simulations, giving rise to so called numerical-kludge waveforms, one of the many prospects to assure that GW templates will be sufficient to faithfully capture true sources and maximise their science yield.

\subsection{Matched filtering and Characteristic Strain}
Given a set of GW waveform templates, how does one go about actually detecting a GW created by an EMRI? We can estimate the strain caused by a typical EMRI, and see that it actually lies orders of magnitude below the expected sensitivity of future space-born detectors: 
\begin{equation}
    h \sim  \frac{f_{\rm z}^{2/3} (G M)^{5/3} q}{c^4 D_{\rm l}} \approx 10^{-23} \frac{q}{10^{-5}} \left( \frac{f_{\rm z}}{10^{-3} \, \rm{Hz}}\right)^{2/3} \left( \frac{M}{10^6 \, \rm{M}_{\odot}}\right)^{5/3} \left( \frac{\rm{Gpc}}{D_{\rm{l}}}\right)
\end{equation}
where we neglected all order one prefactors and also neglect the source's orientation and eccentricity. However, if an accurate waveform template is provided, most of the noise can be extracted out of the datastream via a process referred to as matched filtering, significantly boosting the SNR of the event. The process of matched filtering can be thought of as a particular type of scalar product, in which a data vector $\Vec{D} = \Vec{S} + \Vec{N}$ containing both signal and noise is projected along a filter vector $\Vec{F}$ representing the output of a candidate waveform template. The goal is to find a choice of the filter that will maximise the product $\Vec{S} \cdot \Vec{F}$ while minimising the product  $\Vec{F} \cdot \Vec{N}$. Not surprisingly, the optimal choice for a filter is related to the signal itself, i.e. $\Vec{F} \sim \Vec{S}$, which would represent the ideal case in which the waveform template exactly matches the real GW signal \citep{1998flan,1998flanagan}\footnote{More precisely, the optimal filter is the signal weighted by its loudness over the detector noise.}.
Roughly speaking, an optimal filter (i.e. an accurate waveform template) can boost the SNR of a GW event by a factor proportional to the square root of the number of GW cycles detected at a given frequency. This boosted quantity is often called the characteristic strain, and is a better representation of the potential loudness of of a GW event:
\begin{equation}
    h_{\rm c} \sim \sqrt{T_{\rm{obs}}f_{\rm{z}}} h_0 \sim 3.6\times 10^2 \left(\frac{T_{\rm{obs}}}{4 \, \rm{yr}} \frac{f_{\rm z}}{10^{-3}\, \rm{Hz}} \right)^{1/2} h
\end{equation}
where we assumed a monochromatic source for simplicity.
The frequency of an orbit at the ISCO of a $10^6$ M$_{\odot}$ BH is of the order $2\times 10^{-3}$ Hz, and increases linearly as the mass of the SMBH decreases. This means that under ideal conditions, an EMRI taking place around a relatively light SMBH could reasonably reach characteristic strains on the order of $10^{-20}$, or potentially even $10^{-19}$. These correspond to events that are extremely loud, soaring several orders of magnitudes above the sensitivity of space-born detectors and reaching SNR values of hundreds to thousands. This potential for such high signal to noise ratios from a source orbiting in the strong gravity regime is the key for their incredible scientific potential.

While the SNRs quoted above are expected, it is important to note that in reality the characteristic strain of a realistic EMRI is significantly more complex than the simple circular example shown above, even in the case of simple Newtonian waveforms. In particular, since eccentric orbits slow down at apoapsis and speed up at periapsis, the behaviour of their true anomaly cannot be described by a single frequency. This complexity is reflected in the instantaneous power distribution of eccentric GWs which is spread out over many harmonics rather than being concentrated in a single frequency. Even for the simplest Newtonian waveforms, this spread is described by a complicated sum of Bessel functions $J_{n}$ \citep{1963peters}:
\begin{align}
    h(e) &= h(e=0) \sum_{n=0}^{\infty} g(n,e) \\
    g(n,e) &= \frac{n^4}{32}\big{[}\big{(} J_{n-2}(ne) - 2eJ_{n-1}(ne) + \frac{2}{n}J_{n}(ne)+2eJ_{n+1}(ne)    -J_{n+1}(ne) \nonumber \\&-J_{n+2}(ne)\big{)}^2 + (1-e^2)\big{(}J_{n-2}(ne)-2J_{n}(ne)+J_{n+2}(ne)\big{)}^2 \nonumber \\&+ \frac{4}{3n^2}J_{n}(ne)^2\big{]}.
\end{align}
As the EMRI orbital elements evolve, so do individual harmonic contributions to the total characteristic strain, producing the complex envelopes seen in Figure \ref{fig:15}. Despite all of this complexity (or perhaps because of it), the high SNR achievable by EMRIs is a fantastically enticing prospect for precision GW astronomy, since it can allow to extract the source's parameters such as mass and spin with incredible precision. However, the requirement for a matched filter, i.e. a sufficiently precise GW template is also a double edged sword, since great advances in the study of EMRI waveforms are required in order to exploit this SNR. If all works out, the scientific yield of EMRI waveform will be substantial, as we discuss in Section \ref{sec:potential}.
\begin{figure}
\label{fig:strain}
\centering
\includegraphics[scale=0.7]{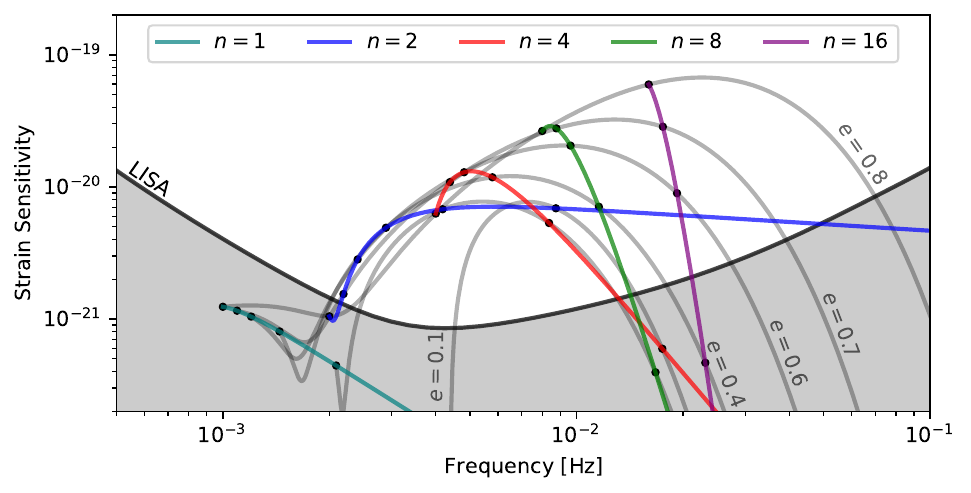} 
\caption{The GW emission of an EMRI evolves in complex ways throughout an observation. Here we plot the sky averaged Newtonian characteristic strain of a plausible eccentric source, initially set to have an eccentricity of $e=0.8$, which evolves from an initial frequency until circularisation (left to right). We highlight the time evolution of several harmonics (coloured lines), while also plotting the instantaneous envelope of all radiated GWs at the specified time-stamps, labelled by the current eccentricity of the EMRI (grey lines). Note how once the source is sufficiently circular, only the $n=2$ harmonic survives above the expected LISA sensitivity (black solid line), while all other harmonics rapidly decay in power.}

\end{figure}

\subsection{Data analysis in practice} 

Now we understand that generating a catalog of all possible EMRI waveforms to adequate precision is challenging. Next we will consider how, even if a complete EMRI waveform catalog exists, how data analysis present its own set of challenges (and how we hope to overcome them).

For analyzing EMRIs in LISA, a few challenges arise with the approach of matched filtering. 
Given the complex nature of EMRI waveforms which are described by $17$ parameters (or more, if environmental perturbations are present), applying the typical matched filtering technique over the vast parameter space of waveforms is not computationally feasible. 
EMRI signals are complex with rich harmonic signatures and they last for several months/years.  While this is a positive aspect in terms of gaining SNR, this makes accurately tracking the waveform over long time scales more challenging (hence the double edged sword). 
As an additional complication, LISA noise is nonstationary, and the data will be susceptible to glitches and gaps that interrupt the buried, long-lived signals. 
Given the vastness of the cosmological volume that LISA will be sensitive to, there may be multiple signals in the LISA data stream simultaneously, which can overlap in time or frequency. This suggests that data analysis of a single source must in some way include knowledge of all other sources that are present in the data. In other words, a global fit of all sources
is needed to account for overlapping signals and to avoid contamination of unmodeled signals \citep{2005PhRvD..72d3005C}.

Luckily several analysis methods are in development to approach these challenges.
Many of these techniques are tested in Mock LISA Data Challenges in preparation for the mission \citep{2010CQGra..27h4009B}. 
One popular approach for GW signal processing is to use Monte-Carlo algorithms (e.g. Markov-Chain: \citealt{Gilks1996,2001PhRvD..64b2001C} or Nested Sampling:  \citealt{2009AIPC.1193..277S,2008arXiv0801.3887C}) to more efficiently explore the parameter space. However due to the multi-harmonic nature of EMRI signals, the parameter space can have multiple regions of high-likelihood which make it difficult to converge. 

Another approach is to perform a search for more generic waveforms that follow the expectations of EMRI orbits (a so-called phenomenological template search, see \citealt{2012PhRvD..86j4050W}), namely that the signal is a superposition of harmonics of three fundamental frequencies, and the phase and amplitude of each harmonic vary slowly in time on a timescale related to the mass ratio.  
Once a signal is detected, one can assume a model to find the best fit to the detected set of harmonics. This approach is ideal for detecting EMRIs subject to unexpected (or poorly modeled) environmental effects, or for testing deviations from GR. In practice, not all harmonics will be recovered equally: stronger, lower frequency harmonics are better recovered overall, and the detection of each harmonic can vary in time as the LISA constellation moves around the Sun.

\section{Science potential with EMRIs}
\label{sec:potential}
\subsection{Fundamental physics}
Observations of EMRIs will provide unique ways to test the limits of our current theory of gravity, GR. Here we focus on a specific example, i.e the notorious ``no-hair theorem", though many more prospects are being investigated.
The no-hair theorem states that the space-time metric around a rotating BH is determined entirely by two parameters: its total mass $M$ and its dimensionless spin $ \xi$. In practice, it states that a spinning BH should be exactly described by the Kerr metric, which is the most general static, asymptotically flat and axisymmetric solution to Einstein's field equations \citep{1963kerr}. Evidence against the no-hairs theorem would be a very strong indication that our current understanding of GR is insufficient, and would likely hint towards a more complete theory of gravity.
The trajectory and the GW emission of an EMRI is largely determined by the exact metric produced by the central SMBH. In particular, the Kerr solution can be expanded into the so called Geroch-Hansen moments, which are a relativistic analog of the spherical harmonic expansion of a Newtonian gravitational potential, a very useful tool to systematically investigate the multipole structure of a metric. For a Kerr black hole, the (complex) Geroch-Hansen moments $\mathcal{M}_{l}$ take a remarkably simple form, which indeed only depends on the BH mass and spin \citep{1974hansen}:
\begin{align}
    \mathcal{M}_{l} = M \left(i \xi \right)^{l}
\end{align}
\noindent where we neglect factors of $c$, $G$ and $\mathcal{M}_l$ vanishes for odd values of $l$.

EMRIs are precise probes of spacetime only a few gravitational radii above the horizon of what are presumably spinning BHs, thus, the detection of a GW will put tight constraints on the moments of the Kerr metric. Particular attention has been devoted to the quadrupole moment $\mathcal{M}_2$, since it enshrines the major difference between a non-spinning BH and a Kerr BH. It is expected that the detection of an EMRI with an SNR of $\sim 100$ will be able to constrain $\mathcal{M}_2$ to a relative precision of $10^{-4}$ \citep{1995ryan,2007curt}! This is very interesting, since many proposed modified gravity theories, i.e. theories that claim to be alternatives to GR, induce deviations the quadrupole moment. We report the explicit form of $\mathcal{M}_2$ of some of the most established alternatives below:
\begin{center}
\begin{tabular}{c|c|c}
  Theory   & $\mathcal{M}_2$ & Further reading \\ \hline 
    GR & $- M \xi^2$ & -- \\
    scalar-tensor & $\frac{m \omega_{\rm{s}}}{3}(1 + \omega_{\rm s})$ & \citet{1913gunnar}\citet{1961brans}\\ 
   f(R) &$ \sqrt{M^2 - q^2}\xi^2$&\citet{1970buchdahl} \\
   EdGB & - M + $\left(\frac{1}{3} + \frac{4 D_1}{3M^2} + \frac{q^2}{12M^2}  \right)$& \citet{1971lovelock} \\
  Kerr-NUT  & $-(M + iN)\xi^2$ & \citet{1976nuts}
\end{tabular}
\end{center}
where the new symbols represent the various free parameters and coupling constants of the modified gravity theories \citep{modgrav,scalaroni,bonnet,2021zi}, e.g. where $\omega_{\rm s}$ corresponds to the frequency (mass) of a presumed massive graviton. Without going into detail, such a varied set of possible modifications to GR would result in a different dynamical evolution of EMRIs and be reflected in their GW emission. Thus, even a single detection of an EMRI is likely to put unprecedented constraints on such theories, deepening our understanding of fundamental physics.

\subsection{Astrophysics with EMRIs }
One of the guaranteed science cases of EMRI detections is to provide precise measurements of SMBH mass and spin. Not only will EMRIs afford better precision compared to electromagnetic probes (e.g. spectral studies of broadened iron lines), but offer the opportunity to study quiescent SMBHs that have typically eluded EM demographic studies. This will provide the opportunity to greatly improve SMBH demographics in the low redshift Universe, which is important for constraining SMBH evolution models. Interested readers can find significant detail in the seminal articles by \citet{2009gair,2010gair}.
For the remainder of this section, however, we will focus on a slightly more speculative, but extremely promising EMRI science prospect, i.e. the detection of so called ``environmental effects".

EMRIs are inevitably born in densely populated, astrophysical environments, which can affect their orbital evolution. This interplay will determine their initial conditions prior to reaching detector bands and alter their trajectories \emph{during} their evolution in the band. In most cases these effects are seemingly negligible, but it must be remembered that the parameter extraction of EMRI signals relies on sufficiently accurate waveform modeling and phase matching. 
If not taken into account, environmental deviations may induce biases in parameter estimation or skew the precision measurements required for tests of GR.  
From an optimistic viewpoint, however, traces of environmental deviations in detected signals present a tantalizing opportunity to learn about EMRI environments, or to independently constrain unknown properties of galactic nuclei. This is particularly exciting for cases where electromagnetic observations are unavailable or limited by resolution.

As discussed in the previous section, 
precise waveforms are required to reach the maximum SNR for a given event. 
Here we discuss how waveform deviations can  occur via astrophysical effects that perturb the EMRI orbit. 
The discussion of environmental effects is intimately tied to tests of GR, since measurements of both require quantifying and analyzing differences in detected waveforms with pre-existing templates derived from GR. In the case of environmental perturbations, the emitted signal will differ from that calculated assuming a source with equivalent parameters is in vacuum.  
In this section we will
 outline one example of an EMRI signal deviation produced by an astrophysical environment, focusing on the case of an EMRI embedded in an AGN accretion disk. 
We leave the reader with a list of other possible environments that may lead to similarly interesting orbital deviations. 
A critical aspect to consider is how various effects evolve during the observed inspiral, focusing in particular on whether they are degenerate with system parameters or other environmental effects, or if they are unique. \\

As discussed in Section~\ref{sec:wetemris}, gas disks can facilitate the formation of EMRIs that coalesce with an SMBH while embedded and co-rotating with the accretion disk. 
For these sources, gas can perturb the orbit of the secondary BH even while it is GW-dominated. 
For a BH on a circular orbit that is co-rotating with the disk, the primary effect of the gas is to exert a gravitational torque. 
Considering only the orbital evolution due to torques, gas effects the orbital evolution of the BH by an amount 
\begin{equation}
\dot{r}_{\rm gas} = 2 T_{\rm gas} \frac{1}{\mu} \frac{1}{\omega r}
\end{equation}
where $\mu$ is the reduced mass of the binary and the gas torque in the linear regime (appropriate for extreme mass ratios) is
\begin{equation}
    T_{\rm gas} \approx \Sigma(r) r^4 \omega^2 q^2 \left(\frac{h}{r}\right)^{-2}
\end{equation}
where again $\Sigma(r)$ is the surface density profile of the accretion disk, $h$ is the disk scale height, and $\omega$ is the orbital frequency. 
For commonly-assumed accretion disk properties (see Section~\ref{sec:wetemris}), one can see that as the binary separation shrinks, the evolution becomes gravitational wave dominated at separations far outside the sensitivity of LISA. 
Once in the LISA band (i.e. the observed GW frequency is $f_{\rm GW}(8 r_{\rm S}(10^6 M_{\odot}),z=1)\sim 0.5$ milliHz), the effect of gas is weak compared to the leading order contribution from GWs for a circular orbit, i.e. $\dot{r}_{\rm gas}/\dot{r}_{\rm GW}\sim 10^{-8} [\Sigma/{10^2 \rm ~g ~cm^{-2}}] [0.01/(h/r)]^2$,  although it can become comparable to higher order PN terms depending on the disk properties \citep{2023MNRAS.521.4645Z}. 

Nevertheless, even weak effects on the inspiral can accumulate to a substantial dephasing over the course of an observation. 
As gas torques affect the inspiral rate of the BH, we can compute the resulting effect on the accumulated phase of the signal. 
The difference in accumulated phase between an event in vacuum ($\phi_{\rm vacuum}$) and an event embedded in gas ($\phi_{\rm gas}$) provides a \emph{phase shift} $\delta \phi$, which for a Newtonian, circular EMRI written in terms of the orbital separation is given by 
\begin{equation}
\begin{split}
 \delta \phi = |\phi_{\rm gas} - \phi_{\rm vacuum}|
    = 2\pi \int_{r_{\rm min}}^{r_{\rm max}} \frac{f}{\dot{r}_{\rm GW}} dr 
      - 2\pi \int_{r_{\rm min}}^{r_{\rm max}} \frac{f}{\dot{r}_{\rm GW} + \dot{r}_{\rm gas}} dr \\
      \approx 
      2 \pi \int_{r_{\rm min}}^{r_{\rm max}} \frac{f \dot{r}_{\rm gas}}{\dot{r}_{\rm GW}^2} \left( 1 + \mathcal{O}\left(\frac{\dot{r}_{\rm gas}}{\dot{r}_{\rm GW}} \right)^2 \right) dr
\end{split}
\end{equation}
where $r_{\min}$ and $r_{\rm max}$ correspond to the rest-frame separation of the binary during the observation window, and $f$ is the rest-frame GW frequency. It is important to remember that the observed GW frequency $f_{\rm obs}$ will be redshifted by $f_{\rm obs} = f/(1+z)$. Here we assume that the impact of gas can be linearly added to the GW component, meaning that the torque mechanism is not affected by GW-driven inspiral (this is not always the case), and in the last expression we assume that the effect of gas is much weaker than GWs (this is likely the case\footnote{We expect that SMBH binaries and most EMRIs in the LISA band will be heavily GW-dominated regardless of the environment. This becomes less certain for stellar-origin BH binaries, whose smaller component masses make their evolution more vulnerable to environmental effects when in the milliHz band.}). 
The phase shift is an observed effect, and so it will depend on the accuracy of our waveform templates as well as the evolutionary stage and observing time of the source. 
For example, we can consider a fiducial EMRI with 
$M=10^6 \rm M_{\odot}$ $q=10^{-5}$ placed at $z=1$ embedded in an accretion disk on a fully circular orbit. This source, if observed during the final year to coalescence, will span a rest frame separation of $\sim 10 r_{\rm S}$ down to the final plunge. The torque exerted by the gas in this region will be sensitive to the disk structure, but we can take a simplifying estimate of a constant surface density and aspect ratio profile, e.g. $\Sigma\sim10 \rm ~g ~cm^{-2}$  and $h/r \sim 0.01$. Taking the above expression, one can see that this gas configuration results in a phase shift of $\delta \phi \sim 0.5$ radians during a $1$ year observation. Considering that this source spans more than $80,000$ orbits during this time span, the magnitude of dephasing may appear insignificant. However, such values are within the range of the accuracy threshold imposed for high SNR source detection (see Section~\ref{sec:GWdyn}) and can be comparable to higher-order PN corrections in the waveform. In practice, the precise value of the phase shift and how it accumulates over the observation (which depends on details of the disk structure and gas morphology) are difficult to predict, given the existing uncertainty in accretion flows on these scales. This means it will be challenging to include accurate estimates of the environment until we improve our numerical models of relativistic, magnetized, accretion flows. Until then, these initial estimates provide motivation for a novel way to probe accretion phenomena with GW detections \citep{2011PhRvL.107q1103Y,2019MNRAS.486.2754D,2021MNRAS.501.3540D,2022arXiv220710086S}.  

Similar dephasing estimates can be estimated for different environments, for example the evolution of an EMRI experiencing gas drag or dynamical friction while evolving through a gas cloud or uniform matter background. Gravitational influence from a substantial dark matter cusp (or spike) can also produce strong dephasing, although the interaction between dark matter and an the inspiraling BH also requires sophisticated modeling \citep{2020PhRvD.102j3022H}. For further reading on the variety of environmental effects, we refer the reader to \citet{2014PhRvD..89j4059B}. 
\\

One may wonder if the accretion of gas during  GW observation would produce a measureable shift in the binary chirp mass. Certainly, such a measurement would provide strong evidence for a matter environment. 
Consider the same fiducial EMRI, this time with both components accreting from an arbitrary source of surrounding gas at their respective Eddington rates, given by $\dot{M}_{\rm Edd} = 4 \pi G M /(\eta \kappa c)$, where we set the opacity to be driven by electron scattering with $\kappa = 0.4\rm ~ cm^2 g^{-1}$, and we take the radiative efficiency to be $\eta = 0.1$. 
During an observation time of $1$ year, the change in the source chirp mass will be of order $\delta\mathcal{M} \sim 10^{-5} \rm M_{\odot}$. With these relatively non-conservative assumptions, 
this shift is still below (albeit approaching) the expected precision of chirp mass measured in parameter estimation \citep{2004barack}. Thus it is accepted that accretion does not play a significant role in altering source parameters during an observation, although this assumption should be reconsidered for sources that may accrete at super-Eddington rates \citep{2020ApJ...892...90C}.

\section{Conclusions: EMRI detection prospects}
In this part of the chapter, we have presented a brief overview of past and current research in the topic of EMRIs. Starting from their astrophysical origin, we then laid-out the major ingredients required for the production of GW templates and the detection of signals.
Compressing this amount of material in a single chapter is challenging, and in many cases depth of knowledge was sacrificed for broader and intuitive understanding of the subject matter. We therefore strongly encourage interested readers to follow up on this chapter with one or more of the comprehensive white papers and reviews that treat EMRIs in much more detail \citep{2018paurev,2019radrec,astrowhite,fundamentalwhite}.
And yet, such a multi-disciplinary view is a vital requirement in order to truly be able to confirm and understand the potential of future detections of EMRI signals. Astrophysical knowledge regarding event rates and typical initial conditions directly factors into the modelling and data analysis pipelines of individual events. Uninformed, or worse yet wrong, priors can significantly hamper the performance of global fitting techniques, often leading to mis-characterised signals or perhaps missing events entirely. Furthermore, after an event is identified, only very careful comparisons between GR, modified gravity and astrophysical effects can lead to confident parameter estimation. In this field, astrophysics, fundamental physics and data analysis are deeply intertwined and substantial progress in one of the former can only be fruitful if it is also reflected in the other. Given that space-born GW detectors will become operative by the mid to late 2030s, this leaves us with approximately a dozen years to understand and tackle crucial open questions, fortify each link that constitutes the long chain of research in EMRIs, and in turn make the most of the truly awesome technology that is planned for GW research.

\bibliographystyle{apj} 
\bibliography{bbois}

\end{document}